\documentclass[prl,twocolumn,aps,superscriptaddress,showpacs,floatfix]{revtex4}

\usepackage{graphicx}
\usepackage{dcolumn}
\usepackage{bm}

\begin{document}
\title{Rectification of spatial disorder}

\author{Jaegon Um}
\affiliation{School of Physics, Korea Institute for Advanced Study,
Seoul 130-722, Korea}
\author{Hyunsuk Hong}
\affiliation{Department of Physics, Chonbuk National University,
Jeonju 561-756, Korea}
\author{Fabio Marchesoni}
\affiliation{Scuola di Scienze e Tecnologie, Universit\`a di Camerino,
I-62032 Camerino, Italy}
\author{Hyunggyu Park}
\affiliation{School of Physics, Korea Institute for Advanced Study,
Seoul 130-722, Korea}
\begin{abstract}
We demonstrate that a large ensemble of noiseless globally coupled-pinned
oscillators is capable of rectifying spatial disorder with spontaneous current
activated through a dynamical phase transition mechanism, either of first or
second order, depending on the profile of the pinning potential.
In the presence of an external weak drive, the same collective mechanism can
result in an absolute negative mobility, which, though not immediately
related to symmetry breaking, is most prominent at the phase transition.
\end{abstract}


 \maketitle

Rectifiers are special devices capable of extracting a steady output signal
(current) even from a perfectly center-symmetric input signal (drive)
\cite{Reimann}. Their minimal operating conditions typically require
an asymmetric internal dynamics, possibly chosen to optimize performance,
and a non-stationary unbiased input signal, mostly a cyclostationary
periodic drive or a time-correlated noise. In thermodynamical terms,
rectifiers are devices that operate under {\it nonequilibrium} conditions.
Direction and magnitude of their output current strongly depend on
both the drive(s) (intensity and time scales) and their intrinsic noise
(temperature). Rectifiers are often called ratchets or else,
to emphasize the role played by fluctuations, Brownian motors~\cite{RMP09}.

Rectification of an external symmetric signal can be strong enough to overcome
the action of an additional static drive, or load, so that the rectifier can
do mechanical work against it. Moreover, under appropriate circumstances,
the load itself induces an asymmetry in the otherwise perfectly symmetric
dynamics of the motor, thus allowing rectification of symmetric ambient
fluctuations in a direction that depends on the details of the modified motor
dynamics. This mechanism can produce the apparently counterintuitive effect of
a spontaneous system response oriented against the load: Rectifiers working
under such operating conditions are said to exhibit absolute negative
mobility (ANM)~\cite{ANM}.

The question now arises whether only time-dependent signals can fuel such
rectification mechanisms. To explore the possibility of replacing temporal
variability by spatial randomness, we considered an ensemble of $N$ coupled
simple phase oscillators, each driven by a static force,
randomly chosen to model their interaction with a spatially homogeneous,
quenched disordered landscape. Their equilibrium phase values coincide with
the local minima of a periodic bistable pinning potential with center-symmetric
cells. Following Kuramoto's prescription~\cite{Kuramoto}, strong cooperative
effects among oscillators are ensured by assuming global pair coupling,
irrespective of spatial separation. At finite temperatures and in the limit
$N\to \infty$, this system is known to undergo spontaneous symmetry breaking
(SSB)~\cite{Buceta}. As a consequence, it generates a spontaneous current in
either direction; moreover, subjected to an external weak global drive, it can also
exhibit ANM and thus work as a motor. Since first detected, both effects have
been interpreted as rectification phenomena induced
by thermal noise~\cite{Buceta,Others}.

\begin{figure}[tp]
\includegraphics*[width=\columnwidth]{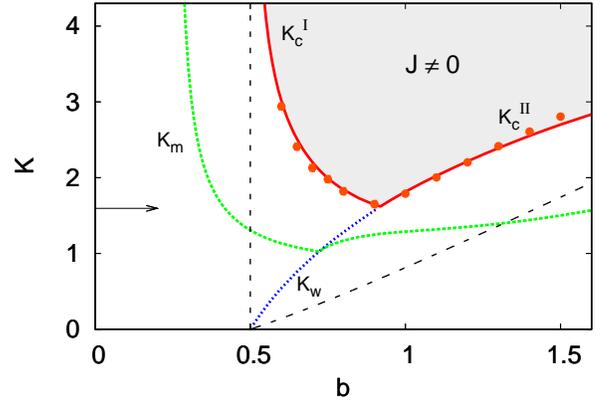}
\caption{(Color online) Phase diagram at $f_e=0$.
The thresholds of  SSB ($K_c$),  global bistability
($K_w$), and  ANM ($K_m$), are plotted versus $b$ when $\sigma_\omega=a=1$.
All curves are obtained by analyzing Eqs.~(\ref{mf_eq})-(\ref{sc_eq})
in the mean field approximation.
Numerical data (dots) are obtained by integrating
Eq.~(\ref{full_eq}) with $N=10^6$, starting from random initial conditions.
Note that the two transition branches ($K_c^I$ and $K_c^{II}$) are connected
by a cusp at $(K^*,b^*)\simeq (1.68,0.94)$.
The dashed curves are the analytic limits of $K_c^I$ and $K_c^{II}$
for $\sigma_\omega = 0$, i.e., $g(\omega)=\delta(\omega)$;
the horizontal arrow indicates the critical coupling for the unpinned Kuramoto model,
$\sqrt{8/\pi} \sigma_\omega$.}
\label{F1}
\end{figure}

In this Letter, we show that a large ensemble of noiseless coupled-pinned
oscillators is capable of rectifying local forces or stresses,
associated with spatially distributed quenched disorder.
By numerically analyzing the stationary dynamics of the oscillator ensemble,
we demonstrate that the onset of spontaneous rectification currents obeys
a dynamical phase transition mechanism, either of first or second order,
depending on the degree of bistability of the pinning potential.
Moreover, we observe that ANM is neither induced by thermal noise,
nor immediately related to symmetry breaking, although its magnitude is
most prominent at the phase transition.

We consider a system of $N$ pinned phase oscillators coupled through
the Kuramoto-type global interaction~\cite{Kuramoto},
\begin{equation} \label{full_eq}
\dot{\phi}_i=\omega_i +f_{e} -V'(\phi_i)
-({K}/{N})\sum_{j=1}^N\sin(\phi_i-\phi_j),
\end{equation}
where $\phi_i$ and $\omega_i$ denote, respectively, the phase and the
intrinsic frequency of the $i$-th oscillator.
The intrinsic frequencies are assumed to be randomly distributed according to
a Gaussian distribution function, $g(\omega)$, with zero mean and
variance $\sigma_\omega$; a tunable frequency bias is introduced by
the external drive $f_e$.
The third and fourth term on the r.h.s. represent the pinning force
acting on the $i$-th oscillator and  all-to-all ferromagnetic ($K>0$) coupling
between oscillators, respectively. Due to the global nature of the interactions,
mean-field (MF)-type phase transitions are expected.

For simplicity, the
pinning potential $V(\phi)$ is chosen to be identical for all oscillators,
\begin{equation} \label{pot_eq}
V(\phi)=-a\cos\phi + ({b}/{2})\cos2\phi,
\end{equation}
with $V'(\phi)=dV/d\phi$.
$V(\phi)$ is a symmetric biharmonic potential, $V(-\phi)=V(\phi)$,
with period $2\pi$. Its unit cells are monostable with one minimum
at $\phi=0$ $({\rm mod}~2\pi)$ for $b\le a/2$, and bistable with two stable
points at $\phi=\pm \phi_m$  for $b>a/2$, where $\phi_m=\arccos(a/2b)$.
Isolated oscillators are
symmetrically locked for $|\omega_i+f_e|<f_p$ with locking phases $|\phi_i|<\phi_p$,
where the largest locking phase $\phi_{p}$ is the (larger) positive solution of
the equation $V''(\phi)=0$ and the depinning frequency threshold is $f_p=V'(\phi_p)$.

In contrast to Refs.~\cite{Buceta,Others}, the coordinates $\phi_i$ are
not coupled to a heat bath. As the ergodic assumption does not apply here,
the oscillator dynamics can depend on the initial conditions (i.c.),
$\{\phi_i(0)\}$. The simulation data reported below refer to the case of
disordered i.c., where $\{\phi_i(0)\}$ have been uniformly randomized
in the interval $(0,2\pi)$. We also simulated ordered initial configurations
with pinned phases, $\phi_i(0)=\phi_m$, and even annealing procedures.
We found that the key qualitative features of our main conclusions do not change.

Following the Kuramoto's approach~\cite{Kuramoto},
we introduce the synchronization order
parameter $\Delta$ and the average phase $\theta$, defined by
$\Delta e^{i\theta} \equiv\langle e^{i\phi_{j}}\rangle$, where
$\langle O_j \rangle=(1/N)\sum_{j=1}^{N}O_j$ denotes the oscillator ensemble average.
Accordingly, Eqs.~(\ref{full_eq}) can be rewritten as
\begin{equation} \label{mf_eq}
\dot{\phi}_i = \omega_i +f_{e} -U'(\phi_i),
\end{equation}
with the {\em effective} potential $U(\phi)$ given by
\begin{equation} \label{mf_pot}
U(\phi)= V(\phi) -K\Delta\cos(\phi-\theta).
\end{equation}
Here, the parameters $\Delta$  and $\theta$
are determined through the the self-consistency relations
\begin{eqnarray}
C&=&\Delta \cos \theta =\langle \cos \phi_j\rangle=\int \cos\left(\phi(\omega)\right)g(\omega)d\omega,\label{sc_eq0}\\
S&=&\Delta \sin \theta =\langle \sin \phi_j\rangle=\int \sin\left(\phi(\omega)\right)g(\omega)d\omega.\label{sc_eq}
\end{eqnarray}

For $\theta=0$, the potential $U(\phi)$ is characterized by
the depinning frequency thresholds $\pm \tilde f_p$, the
corresponding depinning phases $\pm \tilde \phi_p$,
and, if bistable, the two symmetric minima at $\phi=\pm \tilde \phi_m$,
similar to the isolated oscillators.
The  interaction term in $U(\phi)$ tends to suppress the bistability of
the pinning potential even for $b>a/2$:
$U(\phi)$ is bistable only for small $K$, $0 \le K<K_w$,
where the bistability threshold  $K_w$ is the solution of the implicit equation of $K_w \Delta(K_{w})=2b-a$.
For $\theta \neq 0$, $U(\phi)$ is no longer mirror symmetric;
correspondingly, the symmetric depinning thresholds, $\pm \tilde f_p$, are
replaced by two distinct thresholds, $\tilde f_p^{\pm}(K,f_e)$,
with $\tilde f_p^{+} \neq -\tilde f_p^{-}$.

\begin{figure}[tp]
\includegraphics*[width=0.8\columnwidth]{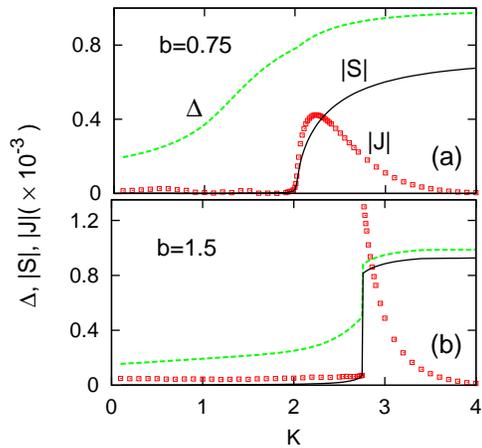}
\caption{(Color online) Spontaneous currents.
$|J|$ (squares), $\Delta$ (dashed) and $|S|$ (solid) versus $K$ (a) in region I
at $b=0.75$, and (b) in region II at $b=1.5$, from the numerical integration of
Eq.~(\ref{full_eq}). }
\label{F3}
\end{figure}

In the absence of external drives ($f_e=0$), the antisymmetry of Eq.~(\ref{mf_eq})
guarantees $\phi(\omega)=-\phi(-\omega)$ as long as $\theta=0$,
so that the current
$J = \langle \dot{\phi} \rangle/2\pi$ is identically zero.
A net SSB current sets on only for $\theta \neq 0$,
where two distinct frequency thresholds ($\tilde f_p^{+} \neq -\tilde f_p^{-}$)
break the left-right symmetry of the running oscillators.
As the pinning ensures a
certain degree of synchronization  ($\Delta >0$ at all $K\ge 0$),
the average phase $\theta$ alone (or more conveniently $S$) becomes the proper order parameter
of the system. Supersymmetry considerations~\cite{Reimann} rule out
spontaneous currents for a purely harmonic pinning potential ($a=0$ or  $b= 0$).
A biharmonic $V(\phi)$, instead, allows nonzero $S$ and $J$ in the strong coupling regime
of $K>K_c$. The transition threshold $K_c$ is actually a function of two parameters only,
$\sigma_\omega$ and $b$, since $a$ can be rescaled to unity without loss of generality.

To investigate a large ensemble of disordered oscillators,
we had recourse to the numerical integration of
the equations of motion, Eq.~(\ref{full_eq}), for the entire ensemble.
For the sake of a comparison, we also numerically solved the self-consistency
equations, Eqs.~(\ref{mf_eq})-(\ref{sc_eq}).
In Fig.~\ref{F1}, we compare the dependence of $K_c$ and $K_{w}$
on the pinning bistability parameter $b$ at $f_e=0$. The intersection of $K_c(b)$
and $K_w(b)$ at $b=b^*$ defines two distinct dynamical regimes: (i)
{\it region I} ($a/2<b<b^*$). The SSB transition occurs after the pinning bistability
is completely suppressed by the Kuramoto coupling ($K_{c}^I>K_{w}$).
The transition branch  $K_c^I$ diverges as $b \to a/2$ as expected
and decays toward a minimum at $b=b^*$;  (ii)
{\it region II} ($b>b^*$). The transition branch $K_c^{II}$ always lies below $K_{w}$~\cite{exp},
so SSB occurs inside the bistability regime where  multiple solutions are possible.
Moreover, we notice that $K_c^I(b)$ and $K_c^{II}(b)$ form a cusp at $b=b^*$, which indicates
that two different SSB mechanisms are at work in regions I and II.

Following this lead, we numerically computed the spontaneous current, $J$,
as a function of $K$ at $b=0.75$ in region I and $b=1.5$ in region II,
see Fig.~\ref{F3}. The difference is revealing: The onset of $J$ can be regarded
as a dynamical phase transition of the {\it second} order, with
\begin{equation} \label{j_eq}
|J| \propto (K-K_c)^{1/2},
\end{equation}
in region I, and of the {\it first} order, in region II. In the latter, $|J|$
jumps discontinuously from $0$ to a maximum at $K=K_c^{II}$, and
finally decays exponentially at larger $K$. Similar continuous or
discontinuous behaviors at the transition points are exhibited by
the corresponding order parameters $\Delta$ and $S$.
No first order transitions were detected in the presence of
noise~\cite{Buceta}.

The existence of a nonzero SSB current for $K>K_c^I$ can be analytically
explained in region I by linearizing Eqs.~(\ref{mf_eq})-(\ref{sc_eq})
for small $S\simeq \Delta \theta$.
A net current can only result from the oscillators running respectively
to the right with $\omega> \tilde f_p^+$ and to the left with
$\omega< \tilde f_p^-$. In the linear regime, it is easy to find
from Eqs.~(\ref{mf_eq}) and (\ref{mf_pot})
$$\tilde f_p^{\pm}\simeq \pm \tilde f_p + K|\cos \tilde \phi_p|S ,$$
which yields the net current $J$ for small drive $f_{e}$
\begin{equation} \label{current}
2\pi J \propto -(\tilde f_p^{+}+\tilde f_p^{-})/2 \simeq  f_e-K|\cos \tilde \phi_p|S.
\end{equation}
Here, Eq.~(\ref{current}) implies a linear response also to
the external drive $f_{e}$, which will be discussed later.
At $f_e=0$, $J \propto -S$ near the transition.

The critical behavior of $S$ near the transition
can be also extracted by linearizing the self-consistency Eqs. (\ref{sc_eq0}) and (\ref{sc_eq})
at $f_e=0$.
Since the system has the Ising-like $Z_2$ symmetry,
Eq.~(\ref{sc_eq}) is expanded in odd powers of $S$ as
\begin{equation}
S=(K/K_c)S-\kappa S^3 + {\cal O}(S^5).
\label{expansion}
\end{equation}
With $\kappa>0$, we find the stable solution of $S=0$ for $K\le K_c^I=K_c$ and
a new stable branch of $S\propto (K-K_c^I)^{1/2}$ for $K\gtrsim K_c^I$,
which determines the MF critical exponent of the current in Eq.~(\ref{j_eq}).

The first order transition along $K_c^{II}(b)$ in region II is a unique feature of our model. It occurs where the static phase solution of
Eq.~(\ref{mf_eq}) for the locked oscillators consists of two {\it disconnected} branches,
$\phi_{1,2}(\omega)$, with $\phi_{1}(-\omega)=-\phi_{2}(\omega)$
in a small $\omega$ interval centered around $\omega=0$.
For a homogeneous randomization of the initial phases,
both solutions $\phi_{1,2}(\omega)$ contribute to the ensemble averages with statistical weights proportional to the respective basin size.
For oscillators with $\omega=0$, these weights are equal at $\theta=0$.
As $\theta$ departs from zero, the symmetry of two branches is broken, but
the deterministic nature of dynamics
does not allow the system to redistribute the oscillator ensemble across
the gap separating two solutions (as it would in the presence
of noise~\cite{Buceta}).
Such a resistance of the system against perturbations
reflects itself in a {\em delayed} onset of the SSB phase in region II; as a consequence, $\kappa$ becomes negative
well before the transition, which causes a discontinuous jump in $\theta$ at the transition.
Indeed, the SSB transition is delayed until the $S=0$ solution becomes {\em locally} unstable
(no matter what the global potential shape is, as the oscillators are unable to redistribute themselves between the two basins).
Therefore, the transition threshold $K_c^{II}$ is again determined by the
linear term in Eq.~(\ref{expansion}), with the difference that for $\kappa <0$ (at the transition)
the stable solution of $S$ exhibits a discontinuity.
This mechanism is distinct from the ordinary first order transition
in equilibrium systems.

In the regime of weak disorder, $\sigma_\omega \ll |\tilde f_p^{\pm}|$,
the contributions from the running (depinned) oscillators to $C$ and $S$ are negligible.
Then, the ensemble averages in Eqs.~(\ref{sc_eq0}) and (\ref{sc_eq}) can be
restricted to static (pinned) oscillators only with $\omega \in (\tilde f_p^{-},\tilde f_p^{+})$.
In this approximation, we can locate $K_c$ in Eq.~(\ref{expansion}) very accurately,
leading to the curves $K_c^I(b)$ and $K_c^{II}(b)$ drawn in  Fig.~\ref{F1}.
One may obtain a simpler but less accurate expression for $K_c$ as
$1/K_c\simeq \langle \cos^2 \phi/U^{\prime\prime} (\phi)\rangle$~\cite{exp1}, which
becomes exact at $\sigma_\omega=0$ for identical oscillators~\cite{exp2}.
Its solutions exhibit, besides the two diverging branches
for $b\to \infty$ and  $b \to a/2$,
also a suggestive lower bound at $b=b^*$ as $\sigma_\omega\sqrt{8/\pi}$,
which happens to coincide with the critical coupling of the
unpinned Kuramoto model [$V(\phi)\equiv 0$].

The reentrant profile of the transition curve $K_c(b)$ can be explained
through a simple qualitative argument. The dynamics of a pinned oscillator in
Eq.~(\ref{full_eq}) results from the competition of two opposing forces:
(i) the random local drive with typical amplitude $\sigma_\omega$;
(ii) the Kuramoto coupling, which is enhanced by a bistable pinning potential.
For a phase transition to occur, the average attractive Kuramoto force must win over
disorder, i.e., $(K/2)\sin2\phi_m=K(a/2b)\sqrt{1-(a/2b)^2}$ must be larger than
$\sigma_\omega$. The non-monotonicity of $\sin 2\phi_m$ in the range
$0\leq \phi_m \leq \pi/2$ determines the convexity of $K_c(b)$.
In particular, $K_c(b)$ diverges in correspondence with the zeros of
$\sin 2\phi_m$, like $\frac{1}{2}(b-a/2)^{-1/2}$ for $b\to a/2$,
and proportional to $2b$ for $b \to \infty$, in agreement with Fig.~\ref{F1}.

\begin{figure}[tp]
\includegraphics*[width=\columnwidth]{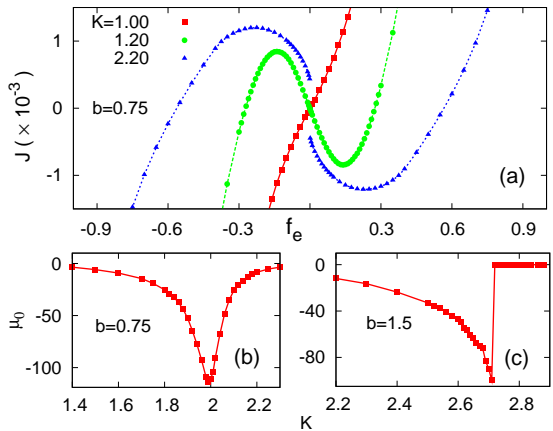}
\caption{(Color online) Absolute negative mobility.
(a) $J$-$f_e$ characteristic curve for $b=0.75$ and three different $K$ from numerical integration of Eq.~(\ref{full_eq}).
(b), (c) $\mu_0$ versus
$K$ in the vicinity of the transition point for (b) $b=0.75$, and (c) $b=1.5$.
For graphical convenience, $\mu_0$ has been rescaled by the factor
$2\pi/ \langle |\dot{\phi}| \rangle$.}
\label{F4}
\end{figure}

Further evidence of the coexisting phase transitions of the first and second
order was obtained by investigating the system response to a finite bias $f_e$,
and more specifically, by analyzing the driven current $J(f_e)$ and its zero-point mobility,
$\mu_0=(dJ/df_e)_{f_e=0}$. We computed both quantities by direct integration
of Eq.~(\ref{full_eq}) and summarized our results in Fig.~\ref{F4}.
In panel (a), the characteristic curve $J$-$f_e$ clearly exhibits three
different regimes: (1) $0<K<K_m$. Below a certain threshold $K_m$ (also plotted versus $b$ in Fig.~\ref{F1}), $J$ is parallel
to $f_e$ as expected in the linear response theory ($\mu_0>0$); (2) $K_m<K<K_c$.
The drive modifies the global interaction of the oscillators with their pinning
potential in such a fashion that the ensemble response
points against the drive. The curve $J(f_e)$ is continuous at $f_e=0$ with
$\mu_0<0$ (ANM), and turns positive for larger $f_e$; (3) $K>K_c$. The slope of
$J(f_e)$ at the origin, $\mu_0$, can grow so negative that eventually the curve
splits into two disconnected antisymmetric branches with $J(0^+)=-J(0^-)<0$ and uniquely
defined negative $\mu_0$ (ANM). This discontinuity is a signature of SSB and is
a common feature of both regions I and II. Note that ANM occurs
regardless of the presence of thermal noise.

The different SSB mechanisms of regions I and II also influence the ANM
properties of the system, see Figs.~\ref{F4} (b) and (c). In region I, $\mu_0(K)$
develops an asymmetric negative peak numerically compatible with a two-sided
divergence for $K\to K_c^{I\pm}$. In region II, $\mu_0(K)$ is clearly
discontinuous with a negative diverging branch for $K\to K_c^{II-}$,
and a slowly decaying one for $K>K_c^{II}$.
In both regions, for exceedingly large or small coupling constants
$\mu_0(K)$ tends to vanish, as expected. Finally, we remark
that ANM does not necessarily anticipate SSB, as apparent in Fig.~\ref{F1},
where the curve $K_m(b)$ (marking the ANM onset) crosses into a region of the monostable pinning for $b<a/2$, inaccessible to $K_c(b)$.

The numerical results of Fig.~\ref{F4} can be qualitatively understood
from the linear response Eq. (\ref{current}) for the current, which
yields
\begin{equation}
\mu_0 \propto 1-K|\cos \tilde \phi_p|\chi,
\label{mobility}
\end{equation}
where $\chi=(dS/df_e)_{f_e=0}$ is the zero-field susceptibility.
The curve $K_m(b)$ plotted in Fig.~\ref{F1} was obtained, indeed,
from Eq.~(\ref{mobility}) and then
checked against the numerical integration data from Eq.~(\ref{full_eq}). Note that
$K_m(b)$ also develops a minimum (actually a cusp) as it crosses $K_w(b)$.
Like in the ordinary equilibrium MF theory, one can find
$\chi \propto |K_c-K|^{-\gamma}$ ($\gamma=1$) near $K=K_c^I$ and, notably, also
for $K\lesssim K_c^{II}$, where the local instability occurs around $S=0$.
Therefore, ANM also diverges at SSB transition point with
MF susceptibility exponent $\gamma=1$, which is consistent with our numerical results.

Finally, among the possible applications of our study, we mention the {\it tug-of-war} 
between two groups of molecular motors~\cite{tugofwar}. Indeed, molecular motors running to the right 
and to the left, due to a Kuramoto type of attraction, can pull polarized filaments in opposite directions. 
However, to treat mobile motors in space, the interaction topology of our model needs to be modified {\it ad hoc}~\cite{mobile}.

This work was supported by Basic Science Research Program
through NRF grant (No.2010-0009697) funded by the MEST. 
We thank KIAS center for Advanced Computation for providing computing resources.


\begin{thebibliography}{6}
\bibitem{exp} In region II, $K_{w}$ is no longer simply given by the equation $K_{w}\Delta =2b-a$,
which is valid only at $\theta= 0$.
\bibitem{exp1} This approximation is good in  regime I for any $b$ and in regime II for $b \to b^*$ and $b \to \infty$.
\bibitem{exp2} For $\sigma_w=0$,
both branches can be obtained analytically (Fig.~\ref{F1}): $K_c^I$ becomes a vertical line at $b=a/2$ and $K_c^{II}=2b+ (a/4b)(a-\sqrt{a^2+32b^2})$.
\end{thebibliography}
\end{document}